# CLIC QD0 "SHORT PROTOTYPE" STATUS

Michele Modena[1]

[1] CERN – TE Department, 1211 Geneva 23 – Switzerland

This paper gives the status design and procurement for the "short prototype" of the QD0 hybrid magnet for CLIC Final Focus system.

## Introduction

The study for the QD0 magnet part of the Final Focus doublets of CLIC Beam Delivery System (BDS) started in 2009.

At that time, the baseline for the Machine Detector Interface (MDI) was defined with an L*=3.5 m and 4.3 m depending by the Detector study proposed. The major consequence of this parameter is that the QD0 magnet will be placed inside the CLIC detectors ILD and SiD as shown in Figure 1 where the QD0 is represented as in the brown element laying on the axis of the cylindrical-symmetric detector

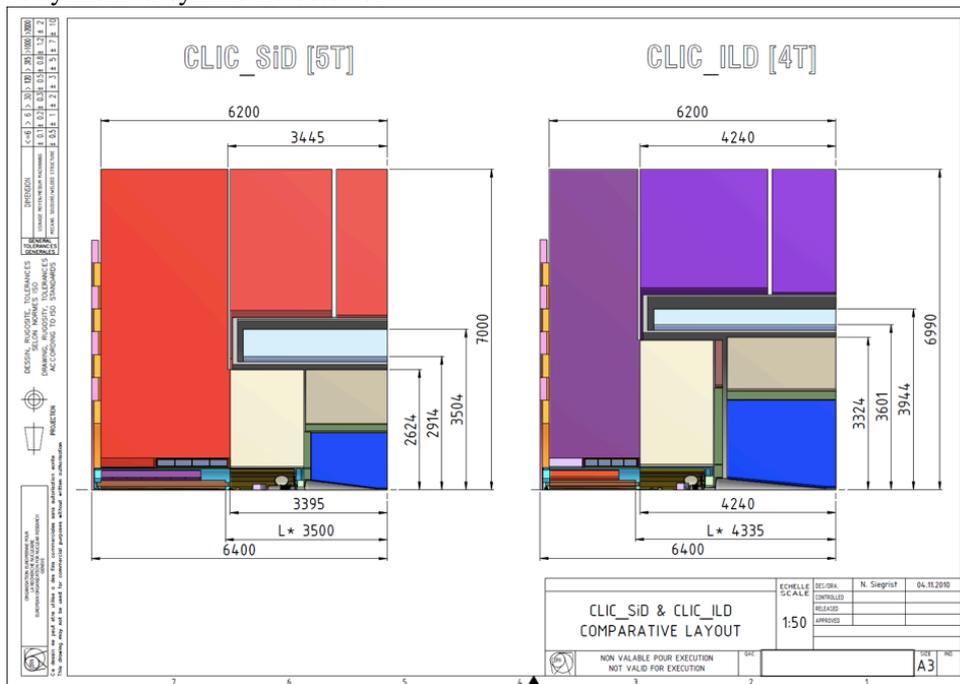

Figure 1: Design Quarter views of the two basic detector layouts of CLIC_SiD and CLIC_ILD

The magnetic requirements for the QD0 (listed in Table 1) are quite severe. To be noted: the extremely high gradient needed, the small aperture of the magnet bore, the length of the magnet, the required tunability.

There are two more conditions which have a strong impact on the design:

- The presence of the post collision line beam pipe beside the QD0: at the L* (front side of the QD0), the distance between the post collision pipeline and the beam axis is barely 35 mm.

-The need of an active stabilization of the quadrupole with the consequences of: a light as possible design; sufficient rigidity and with a well known dynamic behavior (vibration eigenmodes); no source of vibrations (ex. coil coolant flux).

| Parameter | Value |
| --- | --- |
| Nominal field gradient | 575 T/m |
| Nominal integrated field gradient | 1570 T |
| Magnetic length | 2.73 m |
| Magnet bore diameter | 8.25 mm |
| Good field region(GFR) radius | 1 mm |
| Integrated field gradient error inside GFR | < 0.1% |
| Adjustment | +0 to -20% |

Table 1: Magnetic and geometric requirements for the QD0 quadrupole

These two last mentioned boundary conditions seem to exclude a possible superconducting design solution, and the requirement of extremely high gradients, small magnet bore, reduced weight have finally led us to adopt an hybrid electromagnetic/permanent magnet solution.

## QDO Conceptual Design

The proposed cross section of the QDO is shown in Figure 2. The main components: quadrupolar core structure in Permendur, PM blocks, coils, return yokes are visible. A view of the full assembled structure of the Short Prototype of QD0 is also shown in Figure 3.

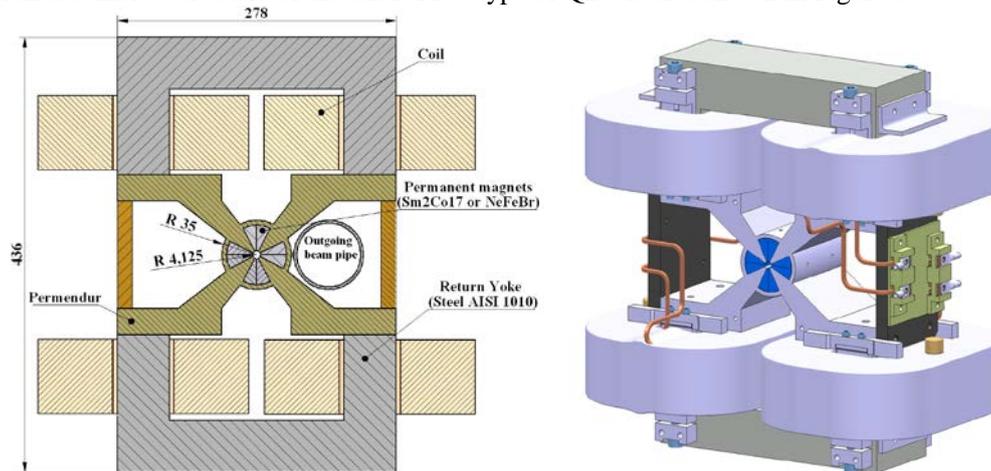

Figure 2-3: Conceptual design of the QD0 cross section and full assembly

LCWS11

A short prototype was built to prove the concept of this quadrupole design. As it is shown in Table 2, all the "cross section parameters" (ex. the gradient, the coil current and current density, the cross-section geometry) are the one needed for the CLIC 3 TeV layout; what is different respect to the "full size" design are the parameters link with the magnet length (ex. voltage, power, and of course mechanical length and weight).

| CLIC QD0 Main Parameters | | 100mm prototype | Real magnet 2.7m |
|---|---|---|---|
| **Yoke** | | | |
| Yoke length | [m] | 0.1 | 2.7 |
| **Coil** | | | |
| Conductor size | [mm] | 4×4 | 4×4 |
| Number of turns per coil | | 18×18=324 | 18×18=324 |
| Average turn length | [m] | 0.586 | 5.786 |
| Total conductor length/magnet | [m] | 0.586×324×4=760 | 5.786×324×4=7500 |
| Total conductor mass/magnet | [kg] | 26.8×4=107.2 | 265.2×4=1060.8 |
| **Electrical parameters** | | | |
| Ampere turns per pole | [A] | 5000 | 5000 |
| Current | [A] | 15.432 | 15.432 |
| Current density | [A/mm$^2$] | 1 | 1 |
| Total resistance | [mOhm] | 896 | 8836 |
| Voltage | [V] | 13.8 | 136.4 |
| Power | [kW] | 0.213 | 2.1 |

Table 2: Magnetic and geometric parameters for the QD0 "Short Prototype" and "Full Size" magnet.

Concerning the magnetic performance of the magnet, simulations were done with: Opera-3D ("OPERA-3D/TOSCA", Vector Fields Limited, Oxford, England).

The hybrid design should guarantee a wide operating range: the achievable gradient versus coils current is shown in Figures 4 and 5. Two different types of permanent magnet material were chosen: $Nd_2Fe_{14}B$ and $Sm_2Co_{17}$. The first compound guarantee the higher performance due to its higher magnetic saturation field, but the second one is the compound that shows higher stability versus radiation and temperature.

In terms of field quality, computations show that the proposed design should fulfill the beam dynamic request. An important aspect concerning the field quality is the design and the manufacturing technology of the central quadrupolar structure in Permendur. In Figures 2 and 3 it can be noted a "ring" structure which solidly connects the four poles. This "ring" structure has no magnetic reasons (at the contrary it slightly decrease the magnet performance in terms of maximum gradient achievable) but has the big advantage to link the poles in a very sound structure also suitable to provide a precise housing for the permanent magnet blocks. The precision of this Permendur structure geometry is guarantee by the chosen machining technology that is EDM (Electric Discharge Machining) wire cutting. The geometric precision of the poles and of the PM blocks positioning is the key aspect to guarantee the required magnetic field quality.

The computed magnetic field components are shown in Figure 6, while Figure 7 shows the Field gradient error (in %) at the boundary of a Good Field Region of 1 mm radius. In both cases data are shown for two types of PM compounds.

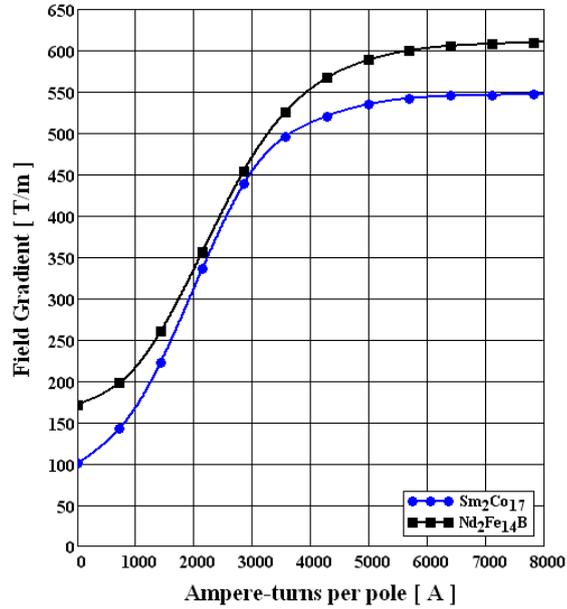

Figure 4: Achievable gradient versus current for two different set of PM blocks compounds.

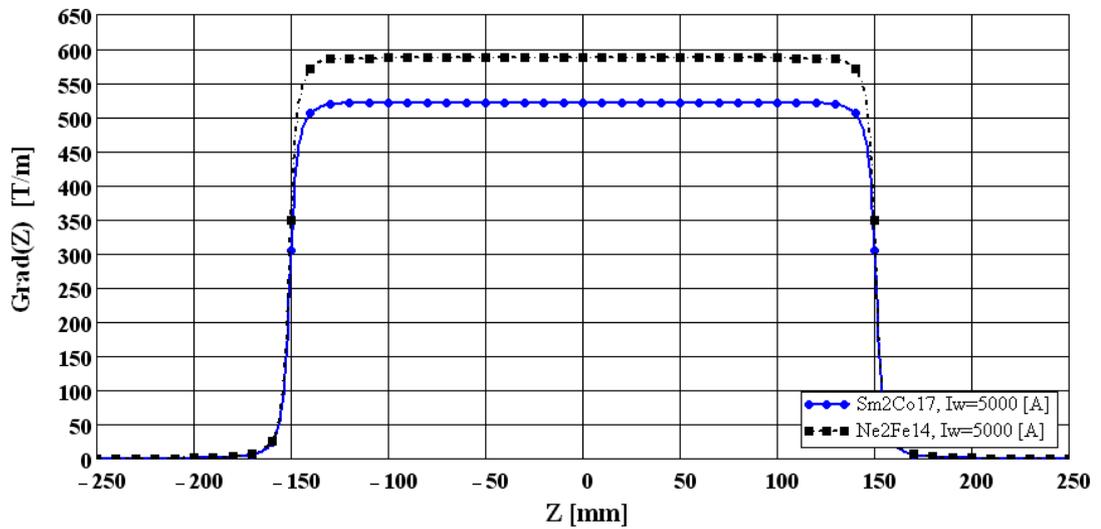

Figure 5: Maximum gradient versus beam axis coordinate for two different set of PM blocks compounds.

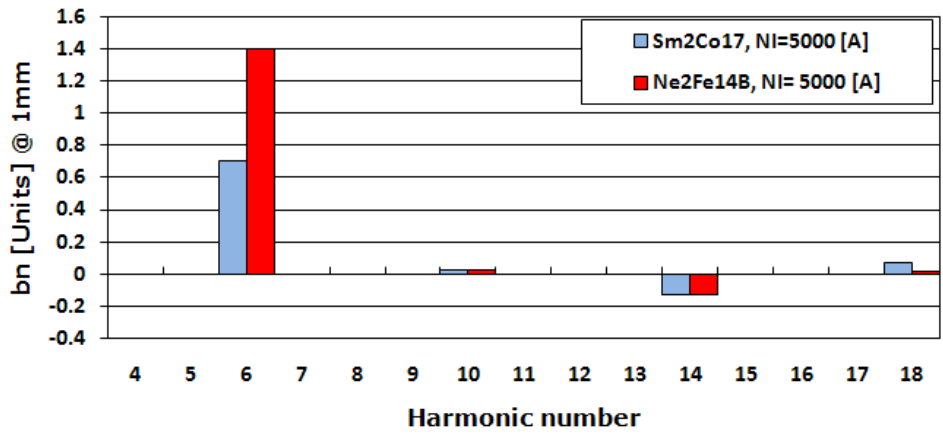

Figure 6: Computed magnetic multipolar components for the QD0 magnet.

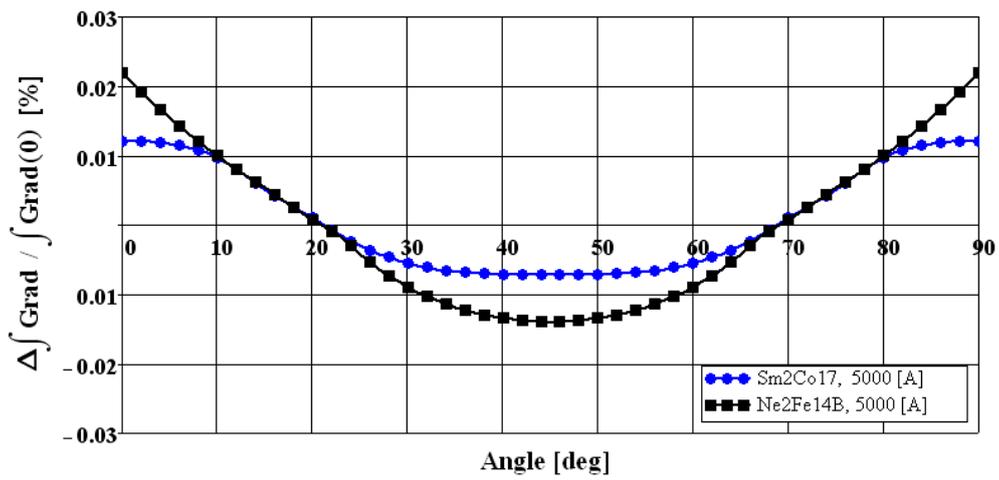

Figure 7: Field quality at a radius of 1 mm (boundary of the required Good Field Region).

## QDO Short Prototype procurement status

The procurement of the components to be produced by EDM technique (the quadrupolar core structure in Pemendur and the PM block sets), was done with Industry (Vacuumschmelze GmbH & Co. KG – Germany) on manufacturing design prepared by CERN. The coils production and the procurement of all other components were done at CERN. Figure 8 shows the completed Permendur part, while a detail of the central part with the PM blocks inserted it is shown in Figure 9.

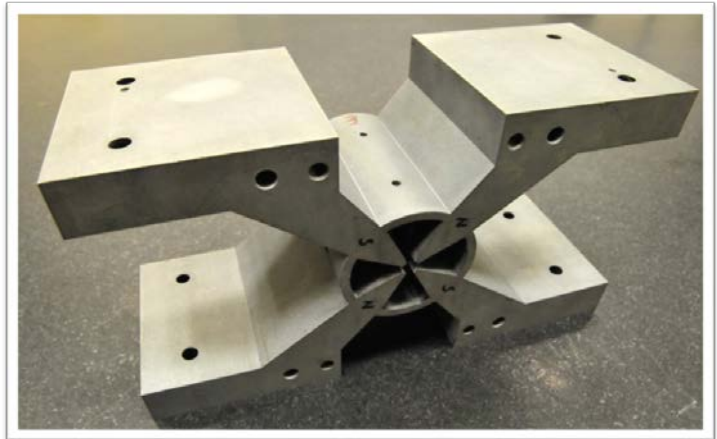

Figure 8: The completed (by EDM technique) magnet core structure in Permendur

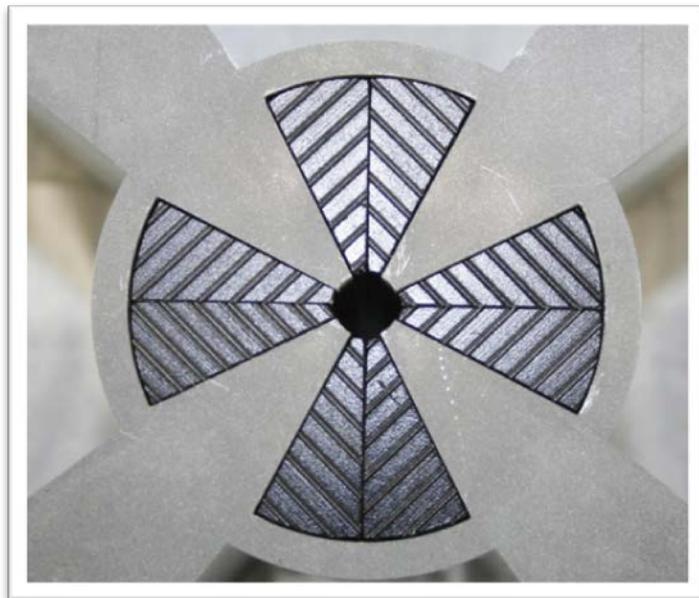

Figure 9: Detail of the magnet central part, the four poles in permendur and the PM blocks are clearly visible.

The mechanical tolerances achieved on the machining of the Permendur part as on the PM blocks are conform to the requirements. As concerning the PM blocks it was not possible to make a measurements of the global magnetization and of its direction tolerances since the blocks are assembled (glued) in set of four and neither Vacuumschmelze nor us were equipped with measuring device (ex. Helmholtz coils) in order to perform the global measurements. Some data on the single blocks magnetic tolerances were anyway made available by the Supplier.

While waiting the procurement of the coils, that were produced with a just commissioned new winding line at CERN, it was possible to perform magnetic measurements of this partially assembled configuration with the PM blocks as the only source of the quadrupolar magnetic field. For this configuration we performed a magnetic simulation in order to compare the measured results. Figures 10 shows a plot of the magnetic computation for this magnet configuration, and Figure 11 shows the magnet on the measuring bench.

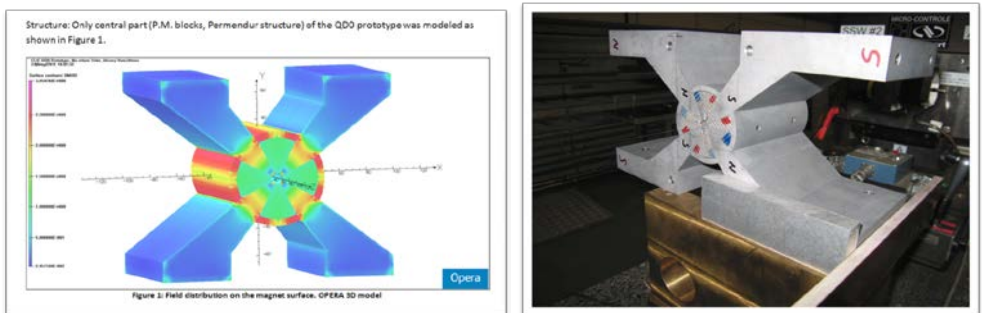

Figure 10-11: FEA computation and magnetic measurements of the partially assembled QD0.

The integrated gradient was measured by SSR (Single Stretched Wire) a very well know technique commonly used for measurement of gradients and magnetic axis, while the multipolar magnetic field components were measured by a new technique called RVW (Rotate Vibrating Wire [2]). The big advantage of method based on stretched wires is to be independent from the magnet aperture.

Results of this first set of magnetic measurements are shown in Figures 12, 13 and 14 for the two types of different PM materials.

| PM material type | Integrated gradient ∫Gdl [T] | |
|---|---|---|
| | MSRD | Calculated |
| VACOMAX 225HR ($Sm_2Co_{17}$) | 15.4 | 15.6 |
| VACODYM 655HR ($Nd_2Fe_{14}B$) | 20.3 | 21.2 |

Figure 12: Computed and measured integrated gradient for the QD0 prototype (for the two different PM compounds).

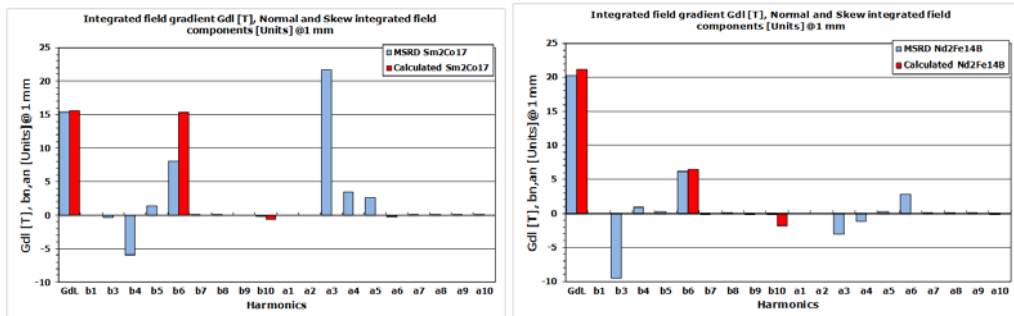

Figure 13-14: Measured field components for the two different PM blocks set.

The visible differences between computed and measured values are probably dependent by the tolerances on the magnetization of the PM blocks (module, directions). The magnetic characterization of the Permendur utilized was done by measurements on bulk material.

The coils were under manufacturing at CERN: at the time of the Granada Workshop, 4 coils were impregnated and were waiting the electrical tests. In Figure 15 it is shown a phase of the coil winding process, while in Figure 16 it is shown a picture of a complete impregnated coil.

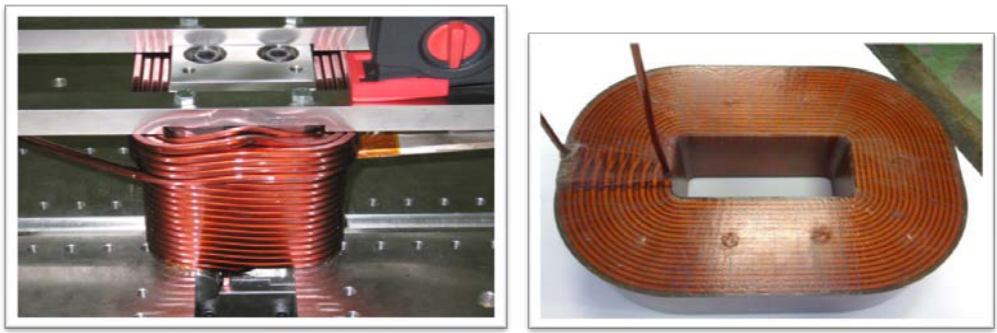

## 4 Future Steps

In the next months we expect to complete the assembly of the prototype and then to perform powering tests and magnetic measurement with the different sets of PM blocks. The main activities planned are:

- Winding and impregnation of 2 spares coils in October 2011.
- Assembly of prototype in its complete configuration in October 2011: all the other ancillary components are available or under finalization at CERN.
- Powering tests and first Magnetic Measurement in November 2011: SSW and RVW system will be utilized.
- Stability tests versus temperature, external magnetic field, radiation, etc: these tests will be planned as soon as the magnetic measurements will be completed.

## Acknowledgements

This work is based on the activities of many person of CERN in particular:
-   The team working (part-time) on QD0 Magnet design and manufacturing:
C. Lopez, E. Solodko, P. Thonet, A. Vorozthsov, and A. Bartalesi
-   The team working on Magnetic Measurements:
M. Buzio, O. Dunkel, J. Garcia Perez, C. Petrone
-   The CERN Metrology Lab that performed acceptance tests on procured components.